\documentclass[reprint, 12pt,a4paper]{elsarticle}
\usepackage{amssymb}
\usepackage{amsmath}
\usepackage{listings}
\usepackage{lineno}
\usepackage{color}
\usepackage[colorlinks=true]{hyperref}
\usepackage{xspace}
\usepackage{multimedia}
\journal{SoftwareX}

\begin{document}

\newcommand{\Aboria}{\href{https://github.com/martinjrobins/Aboria}{Aboria}\xspace}
\newcommand{\GROMACS}{\href{http://www.gromacs.org/}{GROMACS}\xspace}
\newcommand{\Matlab}{\href{https://www.mathworks.com/products/matlab.html}{Matlab}\xspace}
\newcommand{\BLAS}{\href{http://www.netlib.org/blas/}{BLAS}\xspace}
\newcommand{\documentation}{\href{https://martinjrobins.github.io/Aboria/}{online 
documentation}\xspace}
\newcommand{\Eigen}{\href{http://eigen.tuxfamily.org/}{Eigen}\xspace}

\lstset{language=C++,
    frame = single,
    basicstyle=\small\ttfamily,
    keywordstyle=\small\color{blue}\ttfamily,
    stringstyle=\small\color{red}\ttfamily,
    commentstyle=\small\color{magenta}\ttfamily,
    morecomment=\small[l][\color{green}]{\#}
}

\begin{frontmatter}

%% Title, authors and addresses

%% use the tnoteref command within \title for footnotes;
%% use the tnotetext command for theassociated footnote;
%% use the fnref command within \author or \address for footnotes;
%% use the fntext command for theassociated footnote;
%% use the corref command within \author for corresponding author footnotes;
%% use the cortext command for theassociated footnote;
%% use the ead command for the email address,
%% and the form \ead[url] for the home page:
%% \title{Title\tnoteref{label1}}
%% \tnotetext[label1]{}
%% \author{Name\corref{cor1}\fnref{label2}}
%% \ead{email address}
%% \ead[url]{home page}
%% \fntext[label2]{}
%% \cortext[cor1]{}
%% \address{Address\fnref{label3}}
%% \fntext[label3]{}

\title{Particle-based and Meshless Methods with Aboria}

%% use optional labels to link authors explicitly to addresses:
%% \author[label1,label2]{}
%% \address[label1]{}
%% \address[label2]{}

\author[martin]{Martin Robinson}
\author[maria]{Maria Bruna}

\address[martin]{Department of Computer Science, University of Oxford, Wolfson 
    Building, Parks Rd, Oxford OX1 3QD, United Kingdom}
\address[maria]{Mathematical Institute, University of Oxford, Radcliffe 
Observatory Quarter, Woodstock Road, Oxford OX2 6GG, United Kingdom}

\begin{abstract}
    
\Aboria is a powerful and flexible C\texttt{++} library for the implementation 
    of particle-based numerical methods. The particles in such methods can 
    represent actual particles (e.g.  Molecular Dynamics) or abstract particles
    used to discretise a continuous function over a domain (e.g. Radial Basis 
    Functions). Aboria provides a particle container, compatible with the 
    Standard Template Library, spatial search data structures, and a Domain 
    Specific Language to specify non-linear operators on the particle set. This 
    paper gives an overview of Aboria's design, an example of use, and a 
    performance benchmark.

\end{abstract}

\begin{keyword}
%% keywords here, in the form: keyword \sep keyword
particle-based numerical methods\sep
meshless\sep
C++ \sep
%Molecular Dynamics\sep
%Radial Basis Functions
%% PACS codes here, in the form: \PACS code \sep code

%% MSC codes here, in the form: \MSC code \sep code
%% or \MSC[2008] code \sep code (2000 is the default)

\end{keyword}

\end{frontmatter}

%\linenumbers

\section{Motivation and significance}

\Aboria is a C\texttt{++} library that supports the implementation of 
particle-based numerical methods, which we define as having three key 
properties:

\begin{enumerate}

\item There is a set of $N$ particles that have positions within a hypercube of 
    $d$ dimensions, with each dimension being either periodic or non-periodic.
\item The method can be described in terms of non-linear operators on the $N$ particle positions and/or other variables associated to these 
        particles.
\item These operators are defined solely by the particle positions and 
    variables, and typically take the form of interactions 
        between pairs of closely spaced particles (i.e. neighbourhood 
        interactions). There are no pre-defined connections between the 
        particles.

\end{enumerate}

This definition covers a wide variety of popular methods where particles are used to represent either physical particles or a discretisation of a continuous function. In the first category are methods such as Molecular Dynamics 
\citep{griebel2007numerical}, Brownian and Langevin Dynamics \citep{lemons1997}. The second category includes methods like Smoothed Particle Hydrodynamics (SPH) \citep{monaghan2005smoothed}, Radial Basis Functions (RBF) for function interpolation and solution of Partial Differential Equations (PDEs) \citep{hardy1990theory, 
buhmann2003radial, zhang2000meshless}, and Gaussian Processes in Machine Learning \citep{rasmussen2006gaussian}.

To date, a large collection of software has been
developed to implement these methods. Generally, each software package focuses
on one or at most two methods. Molecular and Langevin Dynamics are well-served
by packages such as \GROMACS \citep{abraham2015gromacs},
\href{http://lammps.sandia.gov/}{LAMMPS} \citep{plimpton1995fast},
\href{(http://espressomd.org/}{ESPResSo} \citep{arnold2013espresso} or
\href{https://simtk.org/projects/openmm}{OpenMM} \citep{eastman2012openmm}.
\href{https://wiki.manchester.ac.uk/sphysics/index.php/SPHYSICS_Home_Page}{SPHysics}
\citep{gomez2012sphysics} is one of the best known SPH solvers. There exists no
large-scale package for RBF methods, but these can be implemented in \Matlab
\citep{fasshauer2007meshfree}, and are available as routines in packages such as
\href{https://docs.scipy.org/doc/scipy-0.18.1/reference/generated/scipy.interpolate.Rbf.html}{SciPy} 
\citep{scipy_website}.

The software listed above represents a considerable investment of time and 
money. The computational requirements of particle-based methods, such as the 
efficient calculation of interactions between particles, are challenging to 
implement in a way that scales well with the number of particles $N$, uniform 
and non-uniform particle distributions, different spatial dimensions and 
periodicity. Yet, to date there does not exist a general purpose library that 
can be used to implement these low-level routines, and so they are reimplemented 
again and again in each software package.

The situation with particle-based methods contrasts with mesh-based methods such 
as Finite Difference, Finite Volume or Finite Element Methods. These methods are 
supported by linear algebra libraries based on the \BLAS \citep{blas_website} 
specifications, and today it would be very strange to implement a mesh-based 
method without using BLAS libraries. In general, particle-based methods cannot 
take advantage of linear algebra libraries, which are good for static nodes with 
connections that do not change during the simulation, as opposed to dynamic 
particles whose interaction lists are variable over the course of the 
simulation.

Aboria aims to replicate the success of linear algebra libraries by providing a 
general purpose library that can be used to support the implementation of 
particle-based numerical methods. The idea is to give the user complete control 
to define of operators on the particle set, while implementing 
efficiently the difficult algorithmic aspects of particle-based methods, such as 
neighbourhood searches and fast summation algorithms. However, even at this
level it is not a one-fits-all situation and Aboria is designed to allow users 
to choose specific algorithms that are best suited to the particular 
application. For example, calculating neighbourhood interactions for a uniform 
particle  distribution is best done using a regular cell-list data structure, 
while for a highly non-uniform particle distribution a tree data structure like 
a k-d tree might be preferable \citep{Bentley1979}. For neighbourhood 
interactions that are zero beyond a certain radius, a radial search is the best 
algorithm to obtain interacting particle pairs, while for interactions that are 
amenable to analytical spherical expansions, the fast multipole method is an 
efficient fast summation algorithm \citep{Greengard1997}. 

\subsection{Software capabilities}
\label{sec:capabilities}
    
The \documentation provides a set of example
programs to illustrate Aboria's capabilities. These include:
\begin{description}
  \item[Molecular Dynamics] Simulation of $N$ Newtonian point particles interacting via a linear spring force.

\item[Brownian Dynamics] Simulation of $N$ Brownian point particles moving 
through a set of fixed reflecting spheres that act as obstacles.  

\item[Discrete Element Model (DEM)] Simulation of $N$ granular particles with 
    drag and gravity, interacting via a linear spring force with varying resting 
        lengths. 

\item [Smoothed Particle Hydrodynamics (SPH)] Simulation of a water 
    column in hydrostatic equilibrium. SPH discretises the Navier-Stokes 
        equations using radial interpolation kernels defined over a given 
        particle set. See \cite{robinson2012direct} for more details.

\item[Radial Basis Function (RBF) Interpolation] Computation of an RBF 
    interpolant of a two-dimensional function using a multiquadric basis 
        function. The linear algebra library \Eigen is used to solve the 
        resulting set of linear equations.

\item[Kansa Method for PDEs]  Application of RBFs (using Gaussian basis 
    functions) to solve the Poisson equation in a square two dimensional domain 
        with Dirichlet boundary conditions.

\end{description}

\section{Software description}
\label{sec:description}

\subsection{Software Architecture}

\begin{figure}[htb]
\centering
\includegraphics[width=\textwidth]{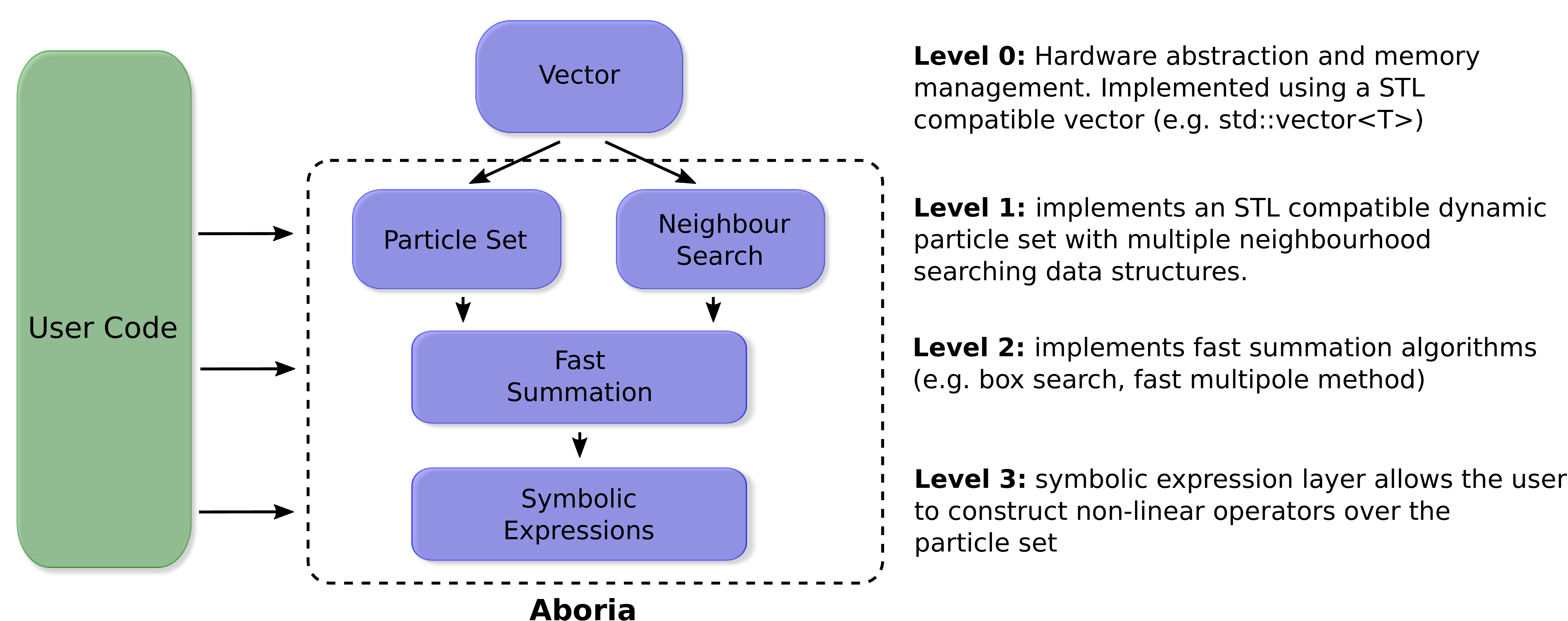}
\caption{\label{fig:aboria_design}{\it Overview of Aboria's design, showing the 
    three different abstraction levels.}}
\end{figure}

The high-level design of Aboria consists of three separate and complimentary 
abstraction levels (see Figure \ref{fig:aboria_design}). Aboria Level 1 contains 
basic data-structures that implement a particle set container, with associated 
spatial search capabilities. Level 2 contains efficient algorithms for 
particle-based methods, such as neighbour searches and fast summation methods.  
Level 3 implements a Domain Specific Language (DSL) for specifying nonlinear 
operators on the set of particles.  While Levels 1 and 2 provide useful 
functionality for particle-based methods, the purpose of Level 3 is to tie 
together this functionality and to provide a easy-to-use interface that ensures 
that the capabilities of Levels 1 and 2 are used in the best possible way.

\subsection{Software Functionalities}
\label{sec:functionalities}

\subsubsection{Aboria Level 1}

\begin{figure}[htb]
\begin{center}
    \includegraphics[width=1\textwidth]{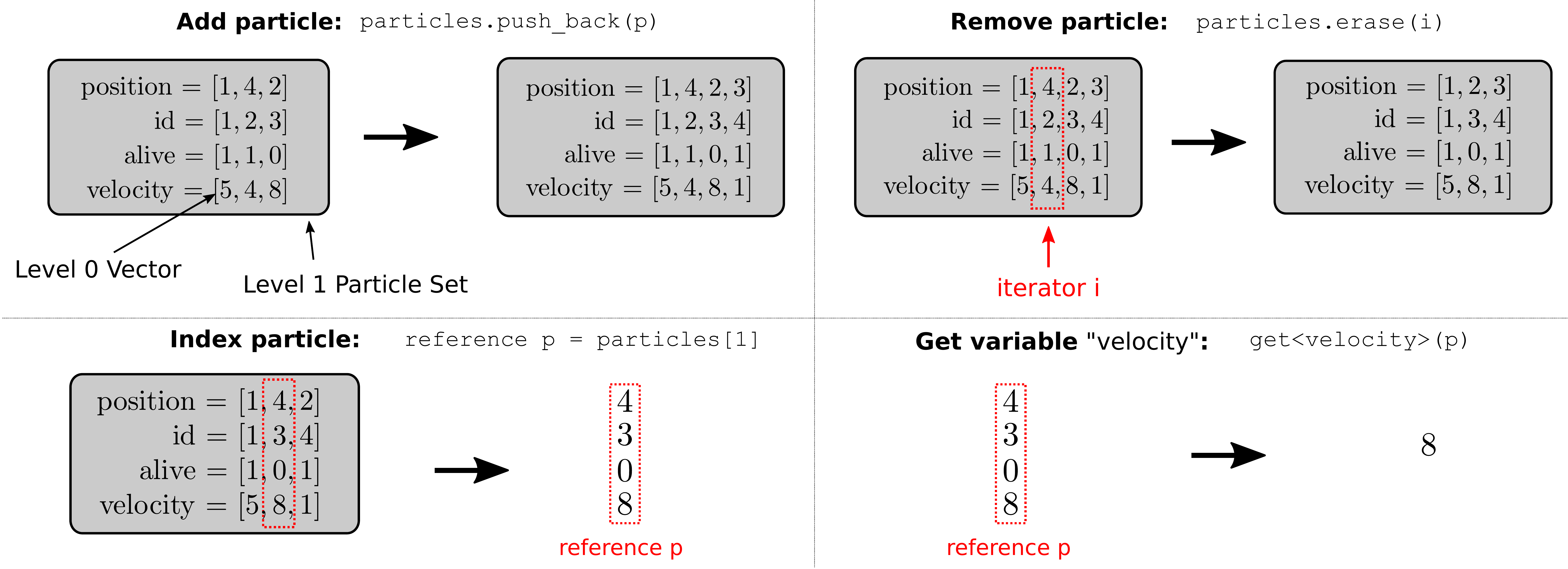}
\end{center}
\caption{\label{fig:particle_set}{\it Operations on an Aboria particle set with 
    $d=1$.  The particle set is displayed as a grey box, with the default 
    variables (\lstinline{position}, \lstinline{id} and \lstinline{alive}) and a 
    user-added variable \lstinline{velocity} shown as horizontal vectors. The 
    data associated to an individual particle correspond to a column in the set 
    (shown as red dashed boxes). (Top left) A new particle \lstinline{p} is 
    added to the    particle set \lstinline{particles}.  (Top right) The 
    particle pointed to by the iterator \lstinline{i} is erased from the 
    particle set.  (Bottom left) Indexing the particle set at index $i$ returns 
    a set of references to the values held at index $i$.  (Bottom right) 
    Individual references to variables can be obtained using \lstinline{get} 
    functions.}}
\end{figure}

Aboria Level 1 implements a particle container class that holds the particle 
data (Figure \ref{fig:particle_set}). This class is based on the Standard Template Library (STL) vector class (Level 0 in 
Figure \ref{fig:aboria_design}), which serves as the lowest-level data 
container. Three variables are stored by 
    default: \lstinline{position}, \lstinline{id} and \lstinline{alive}, 
    representing respectively the particle's spatial position, unique identification number and a flag to indicate if the particle is marked for deletion. The user can specify the spatial  dimension $d$ ($d= 1$ in the example in Figure \ref{fig:particle_set}), as well as any additional variables attached to each particle (an additional variable \lstinline{velocity} is 
    used in Figure \ref{fig:particle_set}). The Level 1 
particle set container will combine multiple Level 0 vectors to form a single 
data structure.

This particle set container generally follows the STL specification
with its own iterators and traits (see Figure \ref{fig:particle_set}). It 
supports operations to add particles (the STL \lstinline{push_back} member 
function), remove particles (\lstinline{erase}), and can return a
single particle given an index $i$ (\lstinline{operator[]}). This index 
operation returns a lightweight type containing references to the corresponding 
index in the set of zipped Level 0 vectors. Individual variables can be obtained 
from this lightweight type via \lstinline{get} functions provided by Aboria.

Level 1 also includes spatial search data structures, which can be used
for fast neighbour searches throughout a hypercube domain with periodic or
non-periodic boundaries. The particle set container interacts with the spatial
search data structures to embed the particles within the domain, ensuring that
the two data structures are kept synchronised while still allowing for updates 
to the particle positions.

\begin{figure}[htb]
\centering
\includegraphics[width=0.8\textwidth]{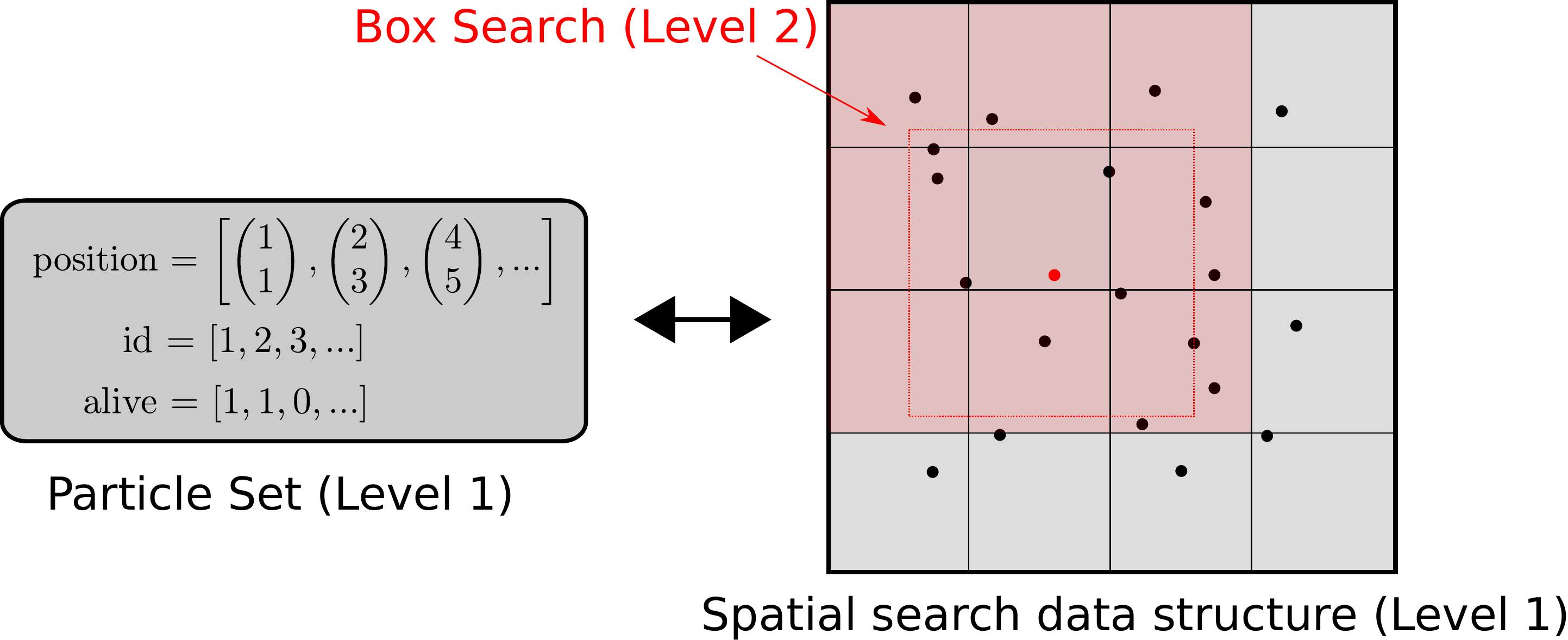}
\caption{\label{fig:Level_1_2} {
 \it Diagram showing the interaction of the two Level 1 data structures in two 
    dimensions ($d=2$). The particle set data structure (left) is kept 
    synchronised with the spatial search data structure 
    (right).  The spatial search data 
    structure is used to implement a box search algorithm in Level 2, which 
    finds all the particles within a box region (red dashed line) around the 
    query point (red point).}}
\end{figure}

The current version of the code implements only 
cell-list search data structures, which divide the domain into a square lattice 
of hypercube cells, each containing zero or more particles (see Figure 
\ref{fig:Level_1_2}). Two cell-list implementations are provided, one which 
supports insertion of particles in parallel by reordering the particles in the 
set, and the other which only supports serial insertion. The latter is faster 
for serial use, and for when particles move rapidly relative to each other (e.g.  
Brownian dynamics). Both data structures support parallel queries. In future 
versions of Aboria a k-d tree data structure will also be added.

\subsubsection{Aboria Level 2}

Aboria Level 2 implements efficient search and fast summation algorithms using 
the particle set and spatial search data structures in Level 1.  Currently Level 
2 includes a box search algorithm around a given spatial position 
$\mathbf{r}_s$.  This calculates which cell $\mathbf{r}_s$ is in and searches 
within that cell and its adjacent cells for potential neighbour
particles within the given search box (see Figure \ref{fig:Level_1_2}). This is 
useful for particle interactions that are zero beyond a certain radius, such as 
the local force interactions of particles in the Molecular Dynamics method. 
Another usage is in the evaluation of compact kernels in the SPH method. 

Currently we are implementing fast multipole methods for summation of global 
dense operators within Aboria, and this level will contain this and any other 
algorithms useful for particle-based methods. 

\subsubsection{Aboria Level 3}
\label{sec:aboria_level_3}

The highest abstraction level in Aboria implements a Domain Specific Language
(DSL) for specifying non-linear operators on the set of particles, using the
\href{http://www.boost.org/doc/libs/1_63_0/doc/html/proto.html}{Boost.Proto} 
library. Users can use standard C\texttt{++} operators (e.g. \lstinline{*}, 
\lstinline{+} or \lstinline{/}) and a
set of supplied functions (e.g. \lstinline{sqrt}, \lstinline{pow} or 
\lstinline{norm}) to express the operator they wish to apply to the given 
particle set.

One key feature of the Aboria DSL is that the symbolic form of each operator is
retained by Aboria, using C\texttt{++} expression template techniques.
Expression templates build up compile-time types that represent the symbolic
form of expressions. For example, the expression \lstinline{a*b + c} in 
C\texttt{++}, where \lstinline{a}, \lstinline{b} and \lstinline{c} are of type 
\lstinline{T}, could be represented using expression templates by the type 
\lstinline{Add<Multiply<T,T>,T>}. In this case, the normal C\texttt{++} 
operators do not calculate a specific value but instead produce a new type that 
encodes which operator was used. 

Like most modern linear algebra libraries (e.g. \Eigen or 
\href{https://bitbucket.org/blaze-lib/blaze}{Blaze}), Aboria uses expression 
templates to enable optimisations and error checking using the symbolic form of 
each operator. For example, Aboria can detect aliasing, which occurs when 
variables are being read and written to within the same operator, and can 
compensate by using a temporary buffer to store the operator results before 
updating the aliased variable. Aboria can detect when a Level 2 fast summation 
method is available, for example when the operator involves a summation over a 
local spatial range, or when the particle positions are altered, thus requiring 
an update of the spatial search data structure. As the symbolic form of the 
operator is known at compile-time, it can be efficiently inlined to all the 
low-level routines implemented in Level 1 and 2, ensuring zero overhead costs 
associated with this level.

Aboria's Level 3 DSL can be used to express vector operations on the 
variables within the particle set, or more complicated non-linear operators that 
involve particle pair interactions, such as neighbourhood force interactions for 
the Molecular Dynamics method, or for the evaluation of RBFs. 

\section{Illustrative Example}
\label{sec:examples}

In this section we describe a complete step-by-step example of how the Aboria library can be used to implement a simple Molecular Dynamics simulation.

We consider $N$ particles within a two-dimensional square domain with 
periodic boundary conditions, interacting via an exponential potential (with cutoff at $r_\text{cut}$). The force on particle $i$ at position $\mathbf{r}_i$ due to particle $j$ at position $\mathbf{r}_j$ is given by 
\begin{equation}
\label{eq:force}
\mathbf{f}_{ij} = \begin{cases}
    -c\, \exp\left({-\|\mathbf{dx}_{ij}\|} \right)
    \frac{\mathbf{dx}_{ij}}{\|\mathbf{dx}_{ij}\|}, & \text{for } 
    \|\mathbf{dx}_{ij}\|<r_\text{cut}, \\
            0, & \text{otherwise},
            \end{cases}
\end{equation}
where $\mathbf{dx}_{ij}=\mathbf{r}_j-\mathbf{r}_i$ is the shortest vector between 
$\mathbf{r}_i$ and $\mathbf{r}_j$ and $c$ is a constant.
We use a semi-implicit Euler integrator with $\delta t=1$ to evolve positions 
$\mathbf{r}_i$ and velocities $\mathbf{v}_i$ with accelerations $\mathbf{a}_i = 
\sum_j \mathbf{f}_{ij}$. This gives the following update equations for each 
timestep $n$ 
\begin{align}
\label{eq:timestep1}
\mathbf{v}^{n+1}_i &= \mathbf{v}^n_i + \sum_j \mathbf{f}^n_{ij} \\
\mathbf{r}^{n+1}_i &= \mathbf{r}^n_i + \mathbf{v}^{n+1}_i.
\label{eq:timestep2}
\end{align}

This is implemented in Aboria as follows. First we define a new type 
\lstinline{velocity} to refer to the two dimensional velocity variable 
$\mathbf{v}$ (using an Aboria \lstinline{double2} type to store the 2D vector)
as well as the particle set type, given by \lstinline{container_type}, which 
contains the velocity variable and has a spatial dimension of $d=2$ (specified 
by the second template argument). For convenience, we also define  
\lstinline{position} as the \lstinline{container_type::position} subclass (we 
will use this later on).  Finally we create \lstinline{particles}, an instance 
of \lstinline{container_type}, containing $N$ particles.

\begin{lstlisting}
ABORIA_VARIABLE(velocity,double2,"velocity")
typedef Particles<std::tuple<velocity>,2> container_type;
typedef typename container_type::position position;

container_type particles(N);
\end{lstlisting}
Next we initialise the positions of the $N$ particles to be uniformly distributed in the unit square (using the standard C\texttt{++} random library), and initialise the 
velocities to zero.
\begin{lstlisting}
std::uniform_real_distribution<double> uni(0,1);
std::default_random_engine g(seed);
for (int i = 0; i < N; ++i) {
    get<position>(particles)[i] = double2(uni(g),uni(g));
    get<velocity>(particles)[i] = double2(0,0);
}
\end{lstlisting}
We then initialise the Level 1 spatial search data structure, providing it with 
lower $(0,0)$ and upper $(1,1)$ bounds for the domain, and setting periodic boundary 
conditions. We set the width of the 
cells in the spatial data structure to \lstinline{r_cut}.
\begin{lstlisting}
particles.init_neighbour_search(double2(0,0),
                                double2(1,1),
                                r_cut,
                                bool2(true,true));
\end{lstlisting}

We now switch to using the Level 3 symbolic API to construct operators over the 
particle set we have created. We define two symbolic objects \lstinline{p} and 
\lstinline{v} representing the position and velocity variables. We also create 
two label objects \lstinline{i} and \lstinline{j} associated to the particle set 
\lstinline{particles}. Note that we define two labels corresponding to indexes $i$ and $j$ symbols in Equation \eqref{eq:force}, so that we can express the interaction force.
\begin{lstlisting}
Symbol<position> p;
Symbol<velocity> v;
Label<0,container_type> i(particles);
Label<1,container_type> j(particles);
\end{lstlisting}
We also create \lstinline{dx}, a symbolic object representing the shortest 
vector between particles $i$ and $j$, and a symbolic accumulation operator 
\lstinline{sum}, using the standard library \lstinline{std::plus} structure to 
perform the accumulation. 
\begin{lstlisting}
auto dx = create_dx(i,j);
Accumulate<std::plus<double2> > sum;
\end{lstlisting}

Now we implement equations \eqref{eq:timestep1} and \eqref{eq:timestep2}. Combining a label with a 
symbol, for example \lstinline{v[i]}, provides us with a concrete vector of 
variables, in this case all the velocity variables in the particle set. 
Similarly \lstinline{p[i]} gives us all the position variables within the 
particle set. 

Using expression templates, as described in Section \ref{sec:aboria_level_3}, we 
can build an  operator involving these objects, as well as the 
\lstinline{dx} and \lstinline{sum} objects described previously. The 
\lstinline{sum} object acts as a summation operator over all the neighbours of
particle $i$. It takes three arguments: (i) a label to accumulate over, (ii) a 
boolean expression that is \lstinline{true} for those particles that will be 
included in the summation, and (iii) an expression that returns the values to be 
added to the summation.  Here we restrict the summation to all particles with 
$\|\mathbf{dx}_{ij}\|<r_\text{cut}$ and $\|\mathbf{dx}_{ij}\|>0$, and sum the 
exponentially decaying force given in Equation \eqref{eq:force}. We then 
accumulate this sum into \lstinline{v[i]}, completing Equation 
\eqref{eq:timestep1}.

\begin{lstlisting}[caption={Velocity update},label={lst:v_update}]
v[i] += c*sum(j, norm(dx)<r_cut && norm(dx)>0,
                    -exp(-norm(dx))*dx/norm(dx)
                    );
\end{lstlisting}

Here, the \lstinline{sum} operator detects that this is a summation 
incorporating only particles within a certain radius $r_\text{cut}$, and uses the Level 
1 spatial search data structure and a Level 2 box search algorithm to speed up 
the summation. In the expression above,  \lstinline{norm} is a symbolic function provided by Aboria that returns the 2-norm of a vector.  

Finally, we write Equation \eqref{eq:timestep2} by incrementing the positions of 
the particles by the value of the velocity.

\begin{lstlisting}[caption={Position update},label={lst:p_update}]
p[i] += v[i];
\end{lstlisting}

\begin{figure}[htb]
\renewcommand\figurename{Video}
\centering
\includegraphics[width=0.6\textwidth]{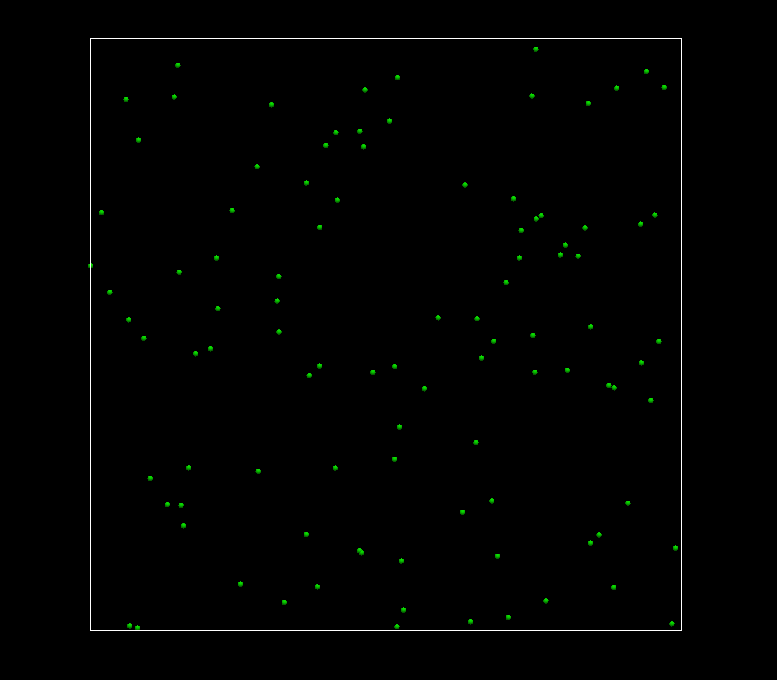}
%\movie[showcontrols]{\includegraphics[width=0.6\textwidth]{example.png}}{example.avi}
\caption{\label{mov:example} {
    {\it Visualisation movie of a Molecular Dynamics simulation of 
    \eqref{eq:timestep1} and \eqref{eq:timestep2} with interaction force 
    \eqref{eq:force}. Each particle position is shown as a green sphere, with a 
    white arrow indicating the direction and magnitude of that particle's 
    velocity. The white outline box shows the simulation domain, which has 
    periodic boundary conditions.}
}}
\end{figure}

Inserting the individual timesteps written in  Listings \ref{lst:v_update} and 
\ref{lst:p_update}  in a time loop completes our implementation of the Molecular 
Dynamics example. The full source code is shown in \ref{sec:code_listing}, and a 
visualisation movie of the positions and velocities of $N=100$ particles over 
$1000$ timesteps is shown in Figure \ref{mov:example}.

To illustrate the efficiency of the Aboria implementation, we benchmark 
the most computationally demanding portion of the program, the update operator 
for \lstinline{v[i]} (Listing~\ref{lst:v_update}). A primary goal for Aboria was zero-cost abstraction: the data structures in Level 1 or the symbolic layer in Level 3 should not incur any additional overhead. 
% If a programmer were to take a Level 0 vector (i.e. a \lstinline{std::vector}), and code by hand a given non-linear operator, this should run at the same speed as the Aboria equivalent.

We compare the Aboria operator given in Listing~\ref{lst:v_update} with a 
hand-coded C\texttt{++} version for a range of particle numbers $N$. Rather than 
re-implementing the neighbour search, we use the search provided in the 
\href{http://manual.gromacs.org/documentation/5.1/doxygen/html-lib/page_analysisnbsearch.xhtml}{analysis 
tools} of the GROMACS library.\footnote{Note that this search algorithm hasn't 
been optimized and is substantially different to the algorithm used by the 
GROMACS simulation engine.} Other than the search facility, the comparison code 
uses standard C\texttt{++} (see the attached supplementary information for the 
benchmark source code).  The execution times $T_e$ are computed as the average 
of $\left \lfloor{10^3/N} \right \rfloor + 1$ evaluations of 
Listing~\ref{lst:v_update}, using a single Intel Core i5-6500T CPU at 2.50GHz 
running Ubuntu 16.04LTS and with 8GB of RAM.

\begin{figure}[htb]
\centering
\includegraphics[width=1.0\textwidth]{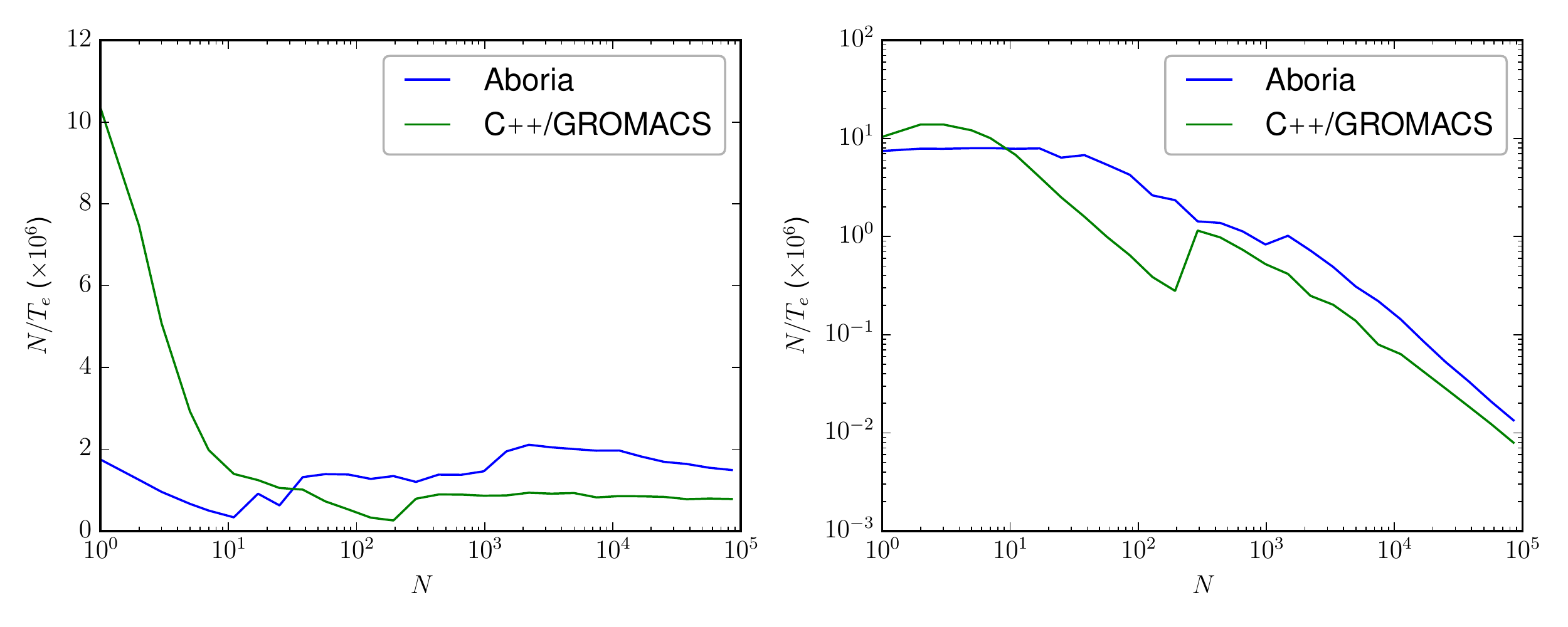}
\caption{\label{fig:example} {
    \it Comparison of the execution time $T_e$ to calculate the velocity update 
    (Equation \eqref{eq:timestep1} and Listing \ref{lst:v_update}) using Aboria 
    versus a hand-coded C++/GROMACS implementation. The scaled inverse execution 
    time $T_e$ is shown versus the number of particles $N$ (a higher value 
    indicates better performance). (Left) Neighborhood search cutoff is $r_{cut}=\sqrt{3/N}$ such 
    that the number of neighbours is fixed. (Right) Cutoff is 
    $r_{cut}=\sqrt{3/500}$ such that the number of neighbours increases with 
    $N$.}}
\end{figure}

In Figure \ref{fig:example} we plot $N/T_e$ versus the number of particles $N$. In the left plot we set $r_\text{cut}=\sqrt{3/N}$ such that the number of neighbors is approximately constant in $N$ (in two dimensions) and hence $T_e$ is expected to scale linearly with $N$. In the right plot we choose a constant cutoff, $r_\text{cut}=\sqrt{3/500}$, so that the number of neighbours increases with 
$N$, and $T_e$ is expected to grow by approximately $N^2$. For very small $N < 
10$ the C\texttt{++}/GROMACS version performs significantly better, as the 
GROMACS search facility turns off the neighbourhood search for small $N$ and 
falls back to a more efficient brute-force $N^2$ method. This is not (yet) 
implemented in Aboria, which tries to do a spatial search, even for very small 
$N$. For $N>10$, the two versions are comparable in performance and, for the 
left hand plot with $r_\text{cut}=\sqrt{3/N}$, exhibit the expected linear 
scaling (i.e.  $N/T_e$ constant). For this two dimensional benchmark, the Aboria 
version is noticeably faster than the C\texttt{++}/GROMACS version for $N>10$, 
however, it should be noted that other benchmarks (described in the 
\documentation) between Aboria and the GROMACS search facility in three 
dimensions put the GROMACS version roughly equal to Aboria.

\section{Impact}

Aboria is designed to make the implementation of particle-based methods easy, 
particularly those involving particle pairwise interactions within local spatial 
neighbourhoods, in any number of dimensions. 
% Its potential impact is analogous to linear algebra libraries, which have already had significant impact since their introduction in the 1970s. For instance, the recent formation and rapid expansion of fields like Data Science or Machine Learning would not be possible without the ubiquitous availability of high quality and efficient linear algebra libraries. 
The design of Aboria was motivated by the lack of a general efficient and high 
quality library for for particle-based methods -- analogous to the linear 
algebra libraries available since the 1970s. With Aboria, new particle-based 
methods or adaptions to existing methods can be proposed and implemented 
quickly, without any sort of performance penalty. Thanks to its ``zero-cost 
abstraction'' design, Aboria can be used as a library within higher-level 
software packages, significantly reducing the implementation effort and 
encouraging the creation of new software that utilises particle-based methods.

Aboria has been used previously in a number of published studies. The core data 
structures and routines were used to implement a coupled SPH-DEM method to 
simulate bidisperse solid particles immersed in a fluid 
\citep{robinson2014fluid}, as well as Smoluchowski Dynamics to simulate a set of 
particles interacting via chemical reactions, with coupling to a lattice-based 
Gillespie method \citep{robinson2014adaptive}. More recently, Aboria was used to 
simulate interacting elliptical particles in a molecular-scale liquid crystal 
model \citep{robinson2017from}, diffusion through random porous media 
\citep{Bruna:2015eh}, and Brownian particles interacting via soft-sphere 
potentials \citep{bruna2017diffusion}.
 
%Aboria is quickly implement different methods to simulate the interacting particles, and we are currently comparing hard- and soft-sphere interactions, with different interaction potentials and integration schemes. 

Until now Aboria has been developed for the use within our research group, which 
has served as a proof of concept for a generic particle-based library. We are 
now expanding to reach other people interested in particle-based models, and 
other application areas. Our research focuses mainly on biological applications, 
in which the use of  particle-based methods has become widespread in recent 
years (see e.g. \cite{osborne2017comparing}). Yet, we are also using Aboria in 
industrial applications projects involving heterogeneous materials where it is 
crucial to account for interactions at the microscale (e.g. membranes and 
batteries). 

\section{Conclusions}

Particle-based numerical methods have already had a large impact on scientific 
progress, primarily through established methods such as Molecular Dynamics, and 
within the chemical physics and materials science communities. However, with 
recent developments in meshless methods, particularly the rise in popularity of 
Radial Basis Functions for interpolation and the solution of PDEs, or related 
developments in Gaussian Processes in Machine Learning, the landscape of 
different particle-based methods has widened significantly. 

Aboria has been created to support the development of these types of numerical 
methods. We hope that this paper will encourage its wider use within the 
community, and that Aboria will have a positive impact by accelerating the 
research and development of particle-based methods, and their application to 
different fields in science and industry.

In the current version of Aboria, we have focussed on establishing the API and 
the three different abstraction levels. Future work will focus on improving 
performance on HPC platforms, and adding new spatial search data structures, 
starting with a k-d tree for non-uniform particle position distributions, and 
new fast summation algorithms, starting with the black-box fast multipole method 
\citep{Fong2009}.

\section*{Acknowledgements}
This work was supported by EPSRC (grant EP/I017909/1), St John's College Research Centre and the John Fell Fund.

%% The Appendices part is started with the command \appendix;
%% appendix sections are then done as normal sections
%% \appendix

%% \section{}
%% \label{}

%% References:
%% If you have bibdatabase file and want bibtex to generate the
%% bibitems, please use
%%
\section*{References}
%\bibliographystyle{elsarticle-num}
%\bibliography{bibliography.bib}

\begin{thebibliography}{10}
\expandafter\ifx\csname url\endcsname\relax
  \def\url#1{\texttt{#1}}\fi
\expandafter\ifx\csname urlprefix\endcsname\relax\def\urlprefix{URL }\fi
\expandafter\ifx\csname href\endcsname\relax
  \def\href#1#2{#2} \def\path#1{#1}\fi

\bibitem{griebel2007numerical}
M.~Griebel, S.~Knapek, G.~Zumbusch, Numerical simulation in molecular dynamics:
  numerics, algorithms, parallelization, applications, Vol.~5, Springer Science
  \& Business Media, 2007.

\bibitem{lemons1997}
D.~S. Lemons, A.~Gythiel, {Paul Langevin's 1908 paper “On the theory of
  Brownian motion” [“Sur la th{\'{e}}orie du mouvement brownien,” C. R.
  Acad. Sci. (Paris) 146, 530–533 (1908)]}, American Journal of Physics
  65~(11) (1997) 1079--1081.

\bibitem{monaghan2005smoothed}
J.~J. Monaghan, Smoothed particle hydrodynamics, Reports on progress in physics
  68~(8) (2005) 1703.

\bibitem{hardy1990theory}
R.~L. Hardy, Theory and applications of the multiquadric-biharmonic method 20
  years of discovery 1968--1988, Computers \& Mathematics with Applications
  19~(8-9) (1990) 163--208.

\bibitem{buhmann2003radial}
M.~D. Buhmann, Radial basis functions: theory and implementations, Cambridge
  Monographs on Applied and Computational Mathematics 12 (2003) 147--165.

\bibitem{zhang2000meshless}
X.~Zhang, K.~Z. Song, M.~W. Lu, X.~Liu, Meshless methods based on collocation
  with radial basis functions, Computational mechanics 26~(4) (2000) 333--343.

\bibitem{rasmussen2006gaussian}
C.~E. Rasmussen, Gaussian processes for machine learning, MIT Press, 2006.

\bibitem{abraham2015gromacs}
M.~J. Abraham, T.~Murtola, R.~Schulz, S.~P{\'a}ll, J.~C. Smith, B.~Hess,
  E.~Lindahl, Gromacs: High performance molecular simulations through
  multi-level parallelism from laptops to supercomputers, SoftwareX 1 (2015)
  19--25.

\bibitem{plimpton1995fast}
S.~Plimpton, Fast parallel algorithms for short-range molecular dynamics,
  Journal of Computational Physics 117~(1) (1995) 1--19.

\bibitem{arnold2013espresso}
A.~Arnold, O.~Lenz, S.~Kesselheim, R.~Weeber, F.~Fahrenberger, D.~Roehm,
  P.~Ko{\v{s}}ovan, C.~Holm, Espresso 3.1: Molecular dynamics software for
  coarse-grained models, in: Meshfree methods for partial differential
  equations VI, Springer, 2013, pp. 1--23.

\bibitem{eastman2012openmm}
P.~Eastman, M.~S. Friedrichs, J.~D. Chodera, R.~J. Radmer, C.~M. Bruns, J.~P.
  Ku, K.~A. Beauchamp, T.~J. Lane, L.-P. Wang, D.~Shukla, et~al., Openmm 4: a
  reusable, extensible, hardware independent library for high performance
  molecular simulation, Journal of Chemical Theory and Computation 9~(1) (2012)
  461--469.

\bibitem{gomez2012sphysics}
M.~Gomez-Gesteira, B.~D. Rogers, A.~J. Crespo, R.~A. Dalrymple,
  M.~Narayanaswamy, J.~M. Dominguez, Sphysics--development of a free-surface
  fluid solver--part 1: Theory and formulations, Computers \& Geosciences 48
  (2012) 289--299.

\bibitem{fasshauer2007meshfree}
G.~E. Fasshauer, Meshfree approximation methods with MATLAB, Vol.~6, World
  Scientific, 2007.

\bibitem{scipy_website}
E.~Jones, T.~Oliphant, P.~Peterson, et~al.,
  \href{http://www.scipy.org/}{{SciPy}: Open source scientific tools for
  {Python}}, [Online; accessed 20/06/2017] (2001--).
\newline\urlprefix\url{http://www.scipy.org/}

\bibitem{blas_website}
B.~T.~B. Forum, \href{http://www.netlib.org/blas/}{Blas (basic linear algebra
  subprograms)}, [Online; accessed 20/06/2017] (1979--).
\newline\urlprefix\url{http://www.netlib.org/blas/}

\bibitem{Bentley1979}
J.~L. Bentley, J.~H. Friedman, {Data structures for range searching}, ACM
  Computing Surveys 11~(4) (1979) 397--409.

\bibitem{Greengard1997}
L.~Greengard, V.~Rokhlin, {A Fast algorithm for particle simulations}, Journal
  of Computational Physics 135~(2) (1997) 280--292.

\bibitem{robinson2012direct}
M.~Robinson, J.~J. Monaghan, Direct numerical simulation of decaying
  two-dimensional turbulence in a no-slip square box using smoothed particle
  hydrodynamics, International Journal for Numerical Methods in Fluids 70~(1)
  (2012) 37--55.

\bibitem{robinson2014fluid}
M.~Robinson, M.~Ramaioli, S.~Luding, {Fluid-particle flow simulations using
  two-way-coupled mesoscale SPH-DEM and validation}, International Journal of
  Multiphase Flow 59 (2014) 121--134.

\bibitem{robinson2014adaptive}
M.~Robinson, M.~Flegg, R.~Erban, Adaptive two-regime method: application to
  front propagation, The Journal of Chemical Physics 140~(12) (2014) 124109.

\bibitem{robinson2017from}
M.~Robinson, C.~Luo, P.~E. Farrell, R.~Erban, A.~Majumdar, From molecular to
  continuum modelling of bistable liquid crystal devices, Liquid Crystals
  (2017) 1--18.

\bibitem{Bruna:2015eh}
M.~Bruna, S.~J. Chapman, {Diffusion in spatially varying porous media}, SIAM J.
  Appl. Math. 75~(4) (2015) 1648--1674.

\bibitem{bruna2017diffusion}
M.~Bruna, S.~J. Chapman, M.~Robinson, Diffusion of particles with short-range
  interactions, Submitted to SIAP.

\bibitem{osborne2017comparing}
J.~M. Osborne, A.~G. Fletcher, J.~M. Pitt-Francis, P.~K. Maini, D.~J. Gavaghan,
  {Comparing individual-based approaches to modelling the self-organization of
  multicellular tissues}, PLOS Computational Biology 13~(2) (2017) 1--34.

\bibitem{Fong2009}
W.~Fong, E.~Darve, {The black-box fast multipole method}, Journal of
  Computational Physics 228~(23) (2009) 8712--8725.

\end{thebibliography}

\appendix
\section{Code listing for Aboria example}
\label{sec:code_listing}

\begin{lstlisting}[frame=none]
void example(const size_t N,
             const unsigned seed, 
             const bool write_out=false) {
    /*
     * set parameters
     */
    const int timesteps = 1e3;
    const double r_cut =  std::sqrt(3.0/N);
    const double c = 1e-3;

    /*
     * Create a 2d particle container type with one 
     * additional variable "velocity", represented 
     * by a 2d double vector
     */
    ABORIA_VARIABLE(velocity,double2,"velocity")
    typedef Particles<std::tuple<velocity>,2> container_type;
    typedef typename container_type::position position;

    /*
     * create a particle set with size N
     */
    container_type particles(N);

    std::uniform_real_distribution<double> uni(0,1);
    std::default_random_engine gen(seed);
    for (int i = 0; i < N; ++i) {
        /*
         * set a random position, and initialise velocity
         */
        get<position>(particles)[i] = double2(uni(gen),uni(gen));
        get<velocity>(particles)[i] = double2(0,0);
    }

    /*
     * initiate neighbour search on a periodic 2d domain 
     * of side length 1
     */
    particles.init_neighbour_search(double2(0,0),
                                    double2(1,1),
                                    r_cut,
                                    bool2(true,true));


    /*
     * create symbols and labels in order to use 
     * the Level 3 API
     */
    Symbol<position> p;
    Symbol<velocity> v;
    Label<0,container_type> i(particles);
    Label<1,container_type> j(particles);

    /*
     * dx is a symbol representing the difference in 
     * positions of particles i and j.
     */
    auto dx = create_dx(i,j);

    /*
     * sum is a symbolic function that sums 
     * a sequence of 2d vectors
     */
    Accumulate<std::plus<double2> > sum;

    /*
     * perform timestepping
     */
    for (int io = 0; io < timesteps; ++io) {

        /*
         * on every step write particle container to a vtk
         * unstructured grid file
         */
        if (write_out) {
            vtkWriteGrid("aboria",io,particles.get_grid(true));
        }

        /*
         * leap frog integrator
         */
        v[i] += c*sum(j, norm(dx)<r_cut && norm(dx)>0,
                    -exp(-norm(dx))*dx/norm(dx)
                    );
        p[i] += v[i];
    }
}
\end{lstlisting}

\section*{Required Metadata}

\section*{Current code version}

\begin{table}[!h]
\begin{tabular}{|l|p{6.5cm}|p{6.5cm}|}
\hline
\textbf{Nr.} & \textbf{Code metadata description} & \textbf{Please fill in this column} \\
\hline
C1 & Current code version & v0.4 \\
\hline
C2 & Permanent link to code/repository used for this code version &  
    \url{https://github.com/martinjrobins/Aboria}\\
\hline
C3 & Legal Code License   & BSD 3-Clause License \\
\hline
C4 & Code versioning system used & git \\
\hline
    C5 & Software code languages, tools, and services used & C\texttt{++} \\
\hline
C6 & Compilation requirements, operating environments \& dependencies & Tested 
    on Ubuntu 14.04LTS with the GCC compiler (version 5.4.1), and Clang compiler 
    (version 3.8.0). Third-party library dependencies: Boost,  Eigen (optional), 
    VTK (optional)\\
\hline
C7 & If available Link to developer documentation/manual & 
    \url{https://martinjrobins.github.io/Aboria} \\
\hline
    C8 & Support email for questions & \url{martin.robinson@cs.ox.ac.uk}\\
\hline
\end{tabular}
\caption{Code metadata (mandatory)}
\end{table}

\end{document}